\documentclass[aps,prl,twocolumn,groupedaddress]{revtex4-2}
\usepackage{aas_macros}
\usepackage{tikz}
\usepackage{amsmath}
\usepackage{bm}
\usepackage{makecell}

\usepackage{xcolor}
\usepackage[version=4]{mhchem}
\begin{document}

%\preprint{}

\title{Multi-species Ion Acceleration in 3D Magnetic Reconnection {with Hybrid-kinetic Simulations}}

 \author{Qile Zhang}
\affiliation{Los Alamos National Laboratory, Los Alamos, NM 87545, USA}
\email[]{qlzhang@lanl.gov}

\author{Fan Guo}
\affiliation{Los Alamos National Laboratory, Los Alamos, NM 87545, USA}
\author{William Daughton}
\affiliation{Los Alamos National Laboratory, Los Alamos, NM 87545, USA}
\author{Hui Li}
\affiliation{Los Alamos National Laboratory, Los Alamos, NM 87545, USA}
\author{Tai Phan}
\affiliation{Physics Department and Space Sciences Laboratory, University of California, Berkeley, Berkeley, CA 94720, USA}
\author{Mihir Desai}
\affiliation{Southwest Research Institute, 6220 Culebra Road, San Antonio, TX 78238, USA}
\affiliation{Department of Physics and Astronomy, University of Texas at San Antonio, San Antonio, TX 78249, USA}
\author{Ari Le}
\affiliation{Los Alamos National Laboratory, Los Alamos, NM 87545, USA}

\date{\today}

\begin{abstract}
\textbf{Magnetic reconnection drives multi-species particle acceleration broadly in space and astrophysics. We perform the first 3D hybrid simulations (fluid electrons, kinetic ions) that contain sufficient scale separation to produce nonthermal heavy-ion acceleration, with fragmented flux ropes critical for accelerating all species. We demonstrate the acceleration of all ion species (up to Fe) into power-law spectra with similar indices, by a common Fermi acceleration mechanism. The upstream ion velocities influence the first Fermi reflection for injection. The subsequent
onsets of Fermi acceleration are delayed for ions with lower charge-mass ratios (Q/M), until growing flux ropes magnetize them. This leads to a species-dependent maximum energy/nucleon $\propto(Q/M)^\alpha$. These findings are consistent with in-situ observations in reconnection regions, suggesting Fermi acceleration as the dominant multi-species ion acceleration mechanism.}

\end{abstract}

%\keywords{}

\maketitle

% body of paper here - Use proper textbf commands

\textbf{Introduction.---}
Magnetic reconnection rapidly converts magnetic energy into bulk flows, heating, and nonthermal particle acceleration. One major unsolved problem is the acceleration of energetic particles during reconnection, with broad implications to various space and astrophysical energetic phenomena \citep{Yamada2010,Matthaeus1988JGR}. Observations have found efficient particle acceleration during reconnection -- with numerous examples from solar flares \citep{Mason2020,Cohen2020}, switchbacks likely from interchange reconnection \citep{Bale2021,Bale2023Nature,Drake2021}, the heliospheric current sheet (HCS) \citep{Desai2022,Phan2022} and the magnetotail \citep{Ergun2018,Ergun2020,Bingham2020,Richard2022}. Often, multiple species are observed, including electrons, protons, and heavier ions \citep{Desai2022,Bingham2020,Richard2022}. These multi-species observations contain key information to discover the underlying acceleration process and can offer more stringent constraints on potential mechanisms. One important candidate is
the Fermi-acceleration mechanism \citep{Drake2006,Dahlin2014,Dahlin2017,Li2017,Li2018,Li2019b,Zhang2021}, where particles get accelerated through curvature drifts along motional electric fields of contracting field lines {\color{black} while bouncing between Alfv\'enic outflows. Other mechanisms have also been proposed, including the Fermi acceleration by bouncing between reconnection inflows \cite{deGouveiadalPino2005AA,Lazarian2012SSRv,Lazarian2020PhPl}, and parallel electric field acceleration \citep{Zhang2019,Zhang2019b,Le2009,Haggerty2015,Dahlin2017,Comisso2018}. }

The recent in-situ observations measure energetic ions near reconnection layers, but the exact energization mechanisms are unknown. Parker Solar Probe {(PSP)} observations near the reconnecting HCS find multi-species energetic ions with maximum energy per nucleon $\varepsilon_{max}\propto(Q/M)^\alpha$ where $\alpha\sim0.65-0.76$ ($M$ is the mass and $Q$ is the charge)\citep{Desai2022}. Some Magnetospheric-Multiscale (MMS) observations at Earth's magnetotail suggest that the ion energization is ordered by energy per charge, which indicates $\alpha \sim 1$ \citep{Bingham2020,Richard2022}. As far as we know, there have not been reconnection theories on multi-species-ion acceleration that can explain these new observations. Drake et al. \citep{Drake2009b,Drake2014pop,Knizhnik2011ApJ} suggested an inverse scaling ($\alpha <0$) in the large-guide-field regime. {For low guide fields, a study of plasma heating \citep{Drake2009} suggested that the temperature is proportional to $M$.} Mechanisms other than reconnection also face significant challenges in explaining HCS observations \cite{Desai2022}.

Fully kinetic simulations have been the primary tools for modeling particle acceleration in collisionless reconnection, as they self-consistently include key reconnection physics and feedback of energetic particles in the reconnection region. 
However, kinetic simulations of reconnection acceleration are still quite challenging due to the multiscale nature of the process. While several large-scale 3D fully kinetic simulations \citep{Zhang2021} have achieved efficient acceleration of electrons and protons, modeling nonthermal acceleration of heavier ions is considerably more difficult due to their large gyroradii ($\propto {(Q/M)^{-1}}$ at the same velocity). Thus, nearly all previous numerical studies on nonthermal acceleration are limited to electrons and/or protons \citep{Zhang2021,Li2019b,Guo_2014,Arnold2021}.
%, where $M_x$ is the mass and $Q_x$ is the charge for individual species. 

Here, {we employ a hybrid (self-consistent particle ions and fluid electrons) model to achieve unprecedentedly large-scale 3D kinetic simulations, to study the acceleration of multi-species ions during reconnection.} Since the hybrid simulations do not need to resolve the electron inertial scale, computationally they are a factor $\sim (d_H/d_e) = (m_H/m_e)^{1/2}$ more cost efficient in each dimension per timestep than fully kinetic simulations. Here $d_H$, $d_e$, $m_H$ and $m_e$ are the inertial lengths and masses of protons and electrons, respectively. Therefore hybrid simulations enable much larger domains to capture the essential physics of heavy ion acceleration. Despite the fluid approximation for electrons, hybrid simulations have demonstrated good agreement for the reconnection rate and dynamics compared to fully kinetic simulations \citep{Stanier2015,Stanier2017,Le2016,Birn2001JGR}. {\color{black}In the Appendix, we show that hybrid and fully kinetic \cite{Zhang2021} simulations produce very similar proton acceleration and flux rope dynamics,
demonstrating that the hybrid model is viable for studying ion acceleration.}

Our hybrid simulations, for the first time, achieved efficient acceleration of multiple ion species ({with a wide range of charge and mass up to \ce{^{56}Fe^{14+}}}) into nonthermal power-law energy spectra. We find that the 3D reconnection layers consist of fragmented kinking flux ropes across different scales {(mainly from the $m=1$ flux-rope kink instability)}, which are growing in both width and length over time, as a distinct component of the reconnection-driven turbulence. {\color{black}The origin and properties of reconnection-driven turbulence are frontiers of research \cite{Beresnyak2017ApJ,Kowal2017ApJ,Huang2016ApJ,Kowal2020ApJ}.} Similar strong and turbulent magnetic fluctuations have also been observed in magnetotail reconnection \citep{Ergun2018,Ergun2020}. This 3D dynamics plays a critical role in the particle acceleration for all species, by facilitating transport to acceleration regions. Different ions are pre-accelerated/injected into nonthermal energies {\color{black}when first bouncing off an Alfv\'enic outflow at a reconnection exhaust (a single Fermi reflection)}. The injection process leads to ``shoulders'' in the energy spectra, becoming the low-energy bounds that control the nonthermal energy content. At higher energy, all species undergo a universal Fermi acceleration process between outflows and form power-law energy spectra with similar indices ($p\sim4.5$).
 However, the onset times of Fermi acceleration {are delayed} for lower charge-mass-ratio ions, until the flux ropes and neighboring exhausts grow large enough to magnetize them. Consequently, the maximum energy per nucleon $\varepsilon_{max}\propto(Q/M)^\alpha$ where $\alpha\sim0.6$ for low upstream plasma $\beta$, and both $p$ and $\alpha$ increase as $\beta$ approaches unity. These results are consistent with the HCS and magnetotail observations \citep{Desai2022,Bingham2020,Richard2022}, suggesting that the observed energetic particles may be a natural consequence of reconnection. 
 
\textbf{Numerical Simulations.---}
We use the Hybrid-VPIC code \citep{Le2021,Bowers2008} that evolves multi-species ions as nonrelativistic particles and electrons as adiabatic fluid, which is coupled with Ohm's law (with small hyper-resistivity and resistivity to break the electron frozen-in condition), Ampere's law and Faraday's law. The simulations start from two identical current sheets (our analyses focus on one) with periodic boundaries and force-free profiles: $B_x=B_0 [\tanh((z-0.25L_z)/\lambda)
-\tanh((z-0.75L_z)/\lambda)-1]$, $B_y=\sqrt{B_0^2+B_g^2-B_x^2}$,
% \begin{equation}
% \begin{split}
% &B_x=B_0 [\tanh((z-0.25L_z)/\lambda)
% -\tanh((z-0.75L_z)/\lambda)-1]\\
% &B_y=[B_0^2 (\sech^2((z-0.25L_z)/\lambda)
% +\sech^2((z-0.75L_z)/\lambda))+B_g^2]^{{1}\over{2}}
% \end{split}
%  \end{equation}
 with uniform density and temperature. We use the initial electron density $n_0$ for the density normalization. $B_0$ is the reconnecting field, $B_g$ is the guide field, $L_z$ is the domain size in $z$, and $\lambda$ is the half thickness of the sheet set to be 1 $d_H$. $b_g=B_g/B_0=0.1$ {(corresponding to a magnetic shear angle $169^{\circ}$), which represents in general the low-guide-field regime in the HCS and magnetotail \citep{Phan2021,Phan2022,Torbert2018,Oieroset2001,Chen2019}}. The domain size $L_x \times L_y \times L_z=1350 \times 140.4 \times 672d_H^3$, with grid size $\Delta x=\Delta y=\Delta z=0.6d_H$ and 800 protons per cell {\color{black} ( $4.7 \times 10^{11}$ protons in total)}. $L_y$ is sufficient for capturing the $m=1$ flux-rope kink mode for efficient acceleration \citep{Zhang2021}. Small long-wavelength perturbations are included to initiate reconnection at both current sheets. To limit the influence of periodic boundaries, the simulations terminate at time $\sim1.3 L_x/V_A$, during which less than 1/3 of the upstream magnetic flux is reconnected and the two current sheets are not yet interacting. We include several ion species \ce{^1H^+,^4He^{2+},^3He^{2+},^{16}O^{7+},^{56}Fe^{14+}}, with abundance 95\%, 5\%, 0.1\%, 0.1\%, 0.1\% respectively. Our simulations are relevant for
multi-X-line collisionless reconnection, as well as
plasmoid reconnection in a thicker current sheet that may develop
kinetic-scale current sheets to trigger collisionless reconnection \citep{Shibata2001,Comisso2017,Uzdensky2016,Daughton2009,Ji2011}.
 
 We present three runs with different initial temperatures $T_i=T_e=0.04, 0.09, 0.25 m_HV_A^2$, where $V_A=B_0/\sqrt{4\pi n_0 m_H}$ is the Alfv\'en speed, resulting in proton $\beta_H=0.08, 0.18, 0.5$ respectively. We discuss the $\beta_H=0.18$ run by default and use others for comparison. Unless otherwise stated, the simulations employ the same initial temperature for all ion species. We have performed additional simulations to confirm that the conclusions are not sensitive to different initial temperatures for different species. 
 
 %To show the insensitivity to this choice, a few additional simulations are considered with different initial species temperatures.} 

\begin{figure*}
\includegraphics[width=1\textwidth]{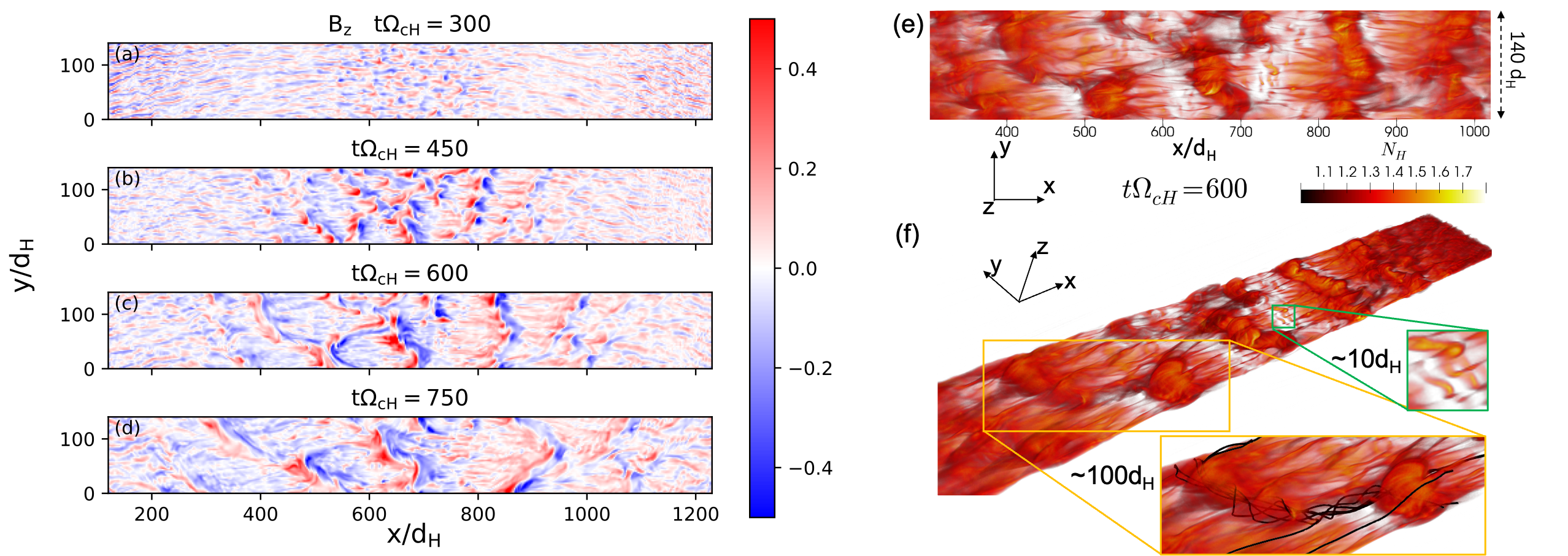}
\caption{%\textbf{Fragmented kinking flux ropes in 3D turbulent reconnection.}
(a)-(d) Magnetic field $B_z$ in the $x-y$ plane in the center of one current sheet ($z=168d_H$) at different times to show the evolution of growing fragmented kinking flux ropes. (e-f) Volume rendering of flux ropes in 3D using proton density at $t\Omega_{cH}=600$ from different perspectives. Two magnified windows ($\sim10d_H$ and $\sim100d_H$) show example kinking flux ropes of vastly different scales in (f). In the $\sim100d_H$ window, several example field lines are shown around the flux rope to better visualize the dynamics.\label{fig1}}
\end{figure*}

\textbf{Reconnection Current Sheet with 3D Fragmented Kinking Flux Ropes.---}
Figure \ref{fig1}(a)-(d) shows $B_z$ in the $x-y$ plane in the center of one current sheet. The unprecedented 3D domain size facilitates strong $m=1$ kink instability of flux ropes that completely fragmentizes the flux ropes -- in contrast to previous smaller-domain simulations \citep{Zhang2021} with more coherent flux ropes {\color{black}(see Appendix)}. This leads to turbulent magnetic fluctuations, as in magnetotail reconnection \citep{Ergun2018,Ergun2020}. As reconnection proceeds, these fragmented kinking flux ropes keep growing over time both in width and length, while they advect along the global bidirectional outflows in $x$. We visualize these flux ropes in 3D in Figure \ref{fig1}(e-f) from different perspectives. Flux ropes in (e) can be directly compared to those in (c), with the same perspective and time. Panel (f) emphasizes that flux ropes exist over a range of scales: one flux rope is newly born from the reconnection layer (green box), and another has grown to occupy a sizeable fraction of the domain (orange box). This flux-rope kink instability produces chaotic field lines \citep{Zhang2021} (see also supplemental material Figure S1) that can diverge quickly and connect outside of the flux ropes \citep{Yang2020,Guo2021}, which enables particles to transport out of flux ropes and get further accelerated at the adjacent reconnection exhausts. 

\begin{figure*}
\includegraphics[width=\textwidth]{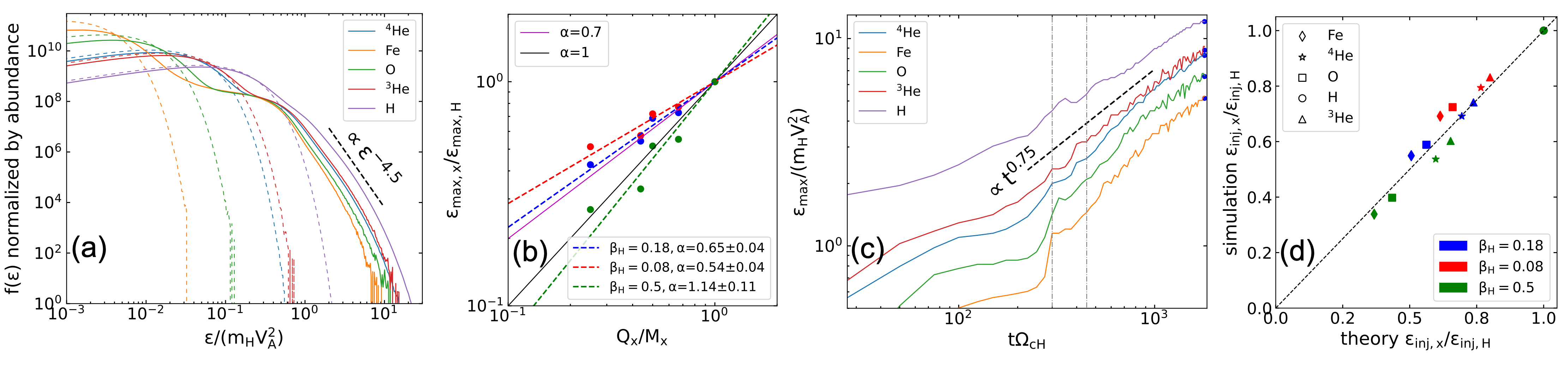}
\caption{%\textbf{Injection and nonthermal acceleration of different ion species.} 
(a) final particle-number spectra (solid lines) normalized by the abundance ratio to Fe, versus energy per nucleon $\varepsilon$. {The initial spectra (dash lines) are also shown for reference.} (b) maximum energy per nucleon $\varepsilon_{max}$ for each species normalized by that of Hydrogen (averaged over time near $t\Omega_{cH}=1800$) versus charge-to-mass ratio, shown for three $\beta_H$ cases. The red, blue and green dash lines fit $(Q_x/M_x)^\alpha$ for different cases, and the fitted $\alpha$ and standard errors are listed in legend. Two solid reference lines $\alpha=0.7,1.0$ indicate different observations \citep{Desai2022,Bingham2020,Richard2022}. (c) $\varepsilon_{max}$ versus time for $\beta_H=0.18$. {Blue dots are marked at $t\Omega_{cH}=1800$ to indicate the values used in (b).} Two vertical grey lines indicate two times evaluated in Figure \ref{fig3}. (d) theoretical estimates of the injection energies (shoulders) normalized by that of Hydrogen, versus those obtained in simulations {at $t\Omega_{cH}=1800$}. \label{fig2}} 
\end{figure*}

\textbf{Acceleration of Different Ions Species.---}
Figure \ref{fig2}(a) shows the particle-number spectra as a function of energy per nucleon $\varepsilon$ for different species at the final time $t\Omega_{cH}=1800$ (solid lines) normalized by their abundance ratio to Fe. For the first time, the simulation shows that all ion species are accelerated into power-law spectra with similar indices $p\sim4.5$, suggesting a universal acceleration process across different ion species. Over time, these nonthermal power laws are formed with sustainable slopes and keep extending to higher energy {(supplemental material Figure S2)}.
%which is similar to the $p\sim-4$ index previously predicted from Fermi acceleration for protons and electrons in the low-beta limit \citep{Zhang2021}.
Moreover, each species develops a shoulder feature in the spectra, marking the low energy bounds of power laws at somewhat different energies. This feature indicates a similar injection process for each species but with intriguing differences, as we will discuss below.
%The low energy bound shoulders of the power laws are all around $m_HV_A^2$ with some difference, suggesting a similar injection process for all species. 
We obtain $\varepsilon_{max}$ as the power-law high-energy cutoffs (where the spectra deviate from the fitted power laws by an e-fold) and show the relative values {near the final time} in Figure \ref{fig2}(b), which follows a fitted scaling $\varepsilon_{max}\propto(Q/M)^\alpha$ ($\alpha\sim0.65$). A simulation with lower $\beta_H=0.08$ produces similar $p\sim4.0$ and $\alpha\sim0.54$, {\color{black}suggesting a low-$\beta$ limit,} while another simulation with higher $\beta_H=0.5$ approaching unity produces $p\sim6.3$ and $\alpha\sim1.14$. 
%range of observed $p\sim4-6$ and $\alpha\sim0.65-1$ near the  In in-situ observations, the upstream $\beta$ is often difficult to precisely measured and variable over time, with typical values 0.1-1 similar to our simulations. Our simulation results agree well with the range of observed $p\sim4-6$ and $\alpha\sim0.65-1$ near the HCS \citep{Desai2022}, and the magnetotail \citep{Bingham2020}. Two reference solid lines $\alpha=0.65,1$ are included in Figure \ref{fig2}(b) to indicate these observations. 
We also performed corresponding 2D simulations and find less efficient acceleration than 3D (supplemental material Figure S3), showing that the 3D dynamics above are critical for particle acceleration of all species.

The time evolution of $\varepsilon_{max}$ (Figure \ref{fig2}(c)) features a common evolution pattern across different species. Different $\varepsilon_{max}$ first increase to $\sim m_HV_A^2$ close to the shoulders in Figure \ref{fig2}(a), entering the nonthermal energies (injection). Later on, different $\varepsilon_{max}$  start increasing at a similar slope roughly following $\varepsilon_{max} \propto t^{0.75}$, once again indicating a universal acceleration process. Simulations with different domain sizes show that the final $\varepsilon_{max}$ is only limited by the acceleration time ($\propto L_x/V_A$). Intriguingly, lower Q/M ions have delayed transitions into the acceleration phase, leading to lower final $\varepsilon_{max}$. Due to the similar acceleration slope, the relative ratios of $\varepsilon_{max}$ between two species will preserve over time and domain sizes, so the ratios in our simulations can extend to larger scales. Our simulation results are consistent with {PSP observations near the HCS} \citep{Desai2022} where upstream $\beta_H\sim 0.2$, {} $\alpha\sim0.65-0.76$ (see Figure \ref{fig2}(b) $\alpha=0.7$ as a reference line) {and $p\sim4-6$ (similar between species considering observational uncertainty)}. In MMS observations {near the magnetotail} \citep{Bingham2020,Richard2022} where upstream $\beta_H$ (usually $<1$) is difficult to measure precisely, the inferred $\alpha\sim1$ and $p\sim5-6$ are comparable to our simulation results (Figure \ref{fig2}(b) $\alpha=1$). 
%The agreement with observations suggests reconnection as the potential source of particle acceleration.

\textbf{Particle Injection and Acceleration Mechanisms.---}
We find that all species are accelerated by a common Fermi acceleration process, with their acceleration rates arising from curvature drifts \citep{Dahlin2014,Li2017} (not shown). We demonstrate the repeated Fermi bounces between outflows with tracer particles in supplemental material (Figure S4(a)). This process produces similar spectral indices $p\sim4.5$ and acceleration $\varepsilon\propto t^{0.75}$ for protons and heavier ions up to {\ce{^{56}Fe^{14+}}}. To our knowledge, this is the first kinetic study demonstrating clear Fermi acceleration of heavier ions. While heavier ions have large gyroradii, the Fermi process can still operate at scales larger than their gyromotion. We also find that a higher initial $\beta$ approaching unity can steepen the power laws by weakening field-line contraction associated with Fermi acceleration. On the one hand, we observed that an initial pressure approaching the magnetic pressure reduces the compression/shrinking at flux ropes related to field-line contraction \citep{Li2018}. On the other hand, this high initial pressure facilitates Fermi acceleration (proportional to parallel energy) that boost the parallel pressure. Therefore, it weakens the firehose parameter $F_h=1-4\pi (P_\parallel-P_\perp)/B^2$ (observed in our simulations) and thus field-line tension ($F_h \textbf{B}\cdot \nabla \textbf{B}/4\pi$) that drives field-line contraction \citep{Arnold2021}. {We have performed additional simulations with different initial temperatures ($0.09-0.25m_HV_A^2$) for the minor ions, and find little changes ($<0.2$) in the spectral slopes. This is because minor ions contribute very little pressure, and can hardly affect the contracting field lines for Fermi acceleration that determines their power-law slopes.}

Before Fermi acceleration, all ion species can be injected through a Fermi reflection when first crossing an exhaust {(supplemental material Figure S4(a-b))}, but are influenced by their initial thermal velocities $V_{th}=\sqrt{T_0/M}$ (lower for heavier ions). {A particle around the initial thermal velocity will get kicked by the exhaust and gain twice of the outflow speed.} Taking the typical outflow speed measured in the simulation (with $\beta_H=0.18$) $V_{out}\sim0.6V_A$, we can roughly estimate the injection energy per nucleon from a single Fermi reflection 
\begin{equation}
    \varepsilon_{inj}\sim0.5m_H(2V_{th}+2V_{out})^2=2m_H(V_{th}+V_{out})^2. \label{inj_en}
\end{equation}
{We have used initial velocity $2V_{th}$ near the higher-energy drop-off of the initial Maxwellian energy spectra, which will approximately correspond to the shoulder after the Fermi reflection.} This theoretical estimate agrees approximately with the shoulders in Figure \ref{fig2}(a) (determined at a level $10^7$), as demonstrated in Figure \ref{fig2}(d).
%and the first increase of $\varepsilon_{max}$ in Figure \ref{fig2}(c) (will show an extra figure in the full paper). 
%here

\begin{figure*}
\includegraphics[width=\textwidth]{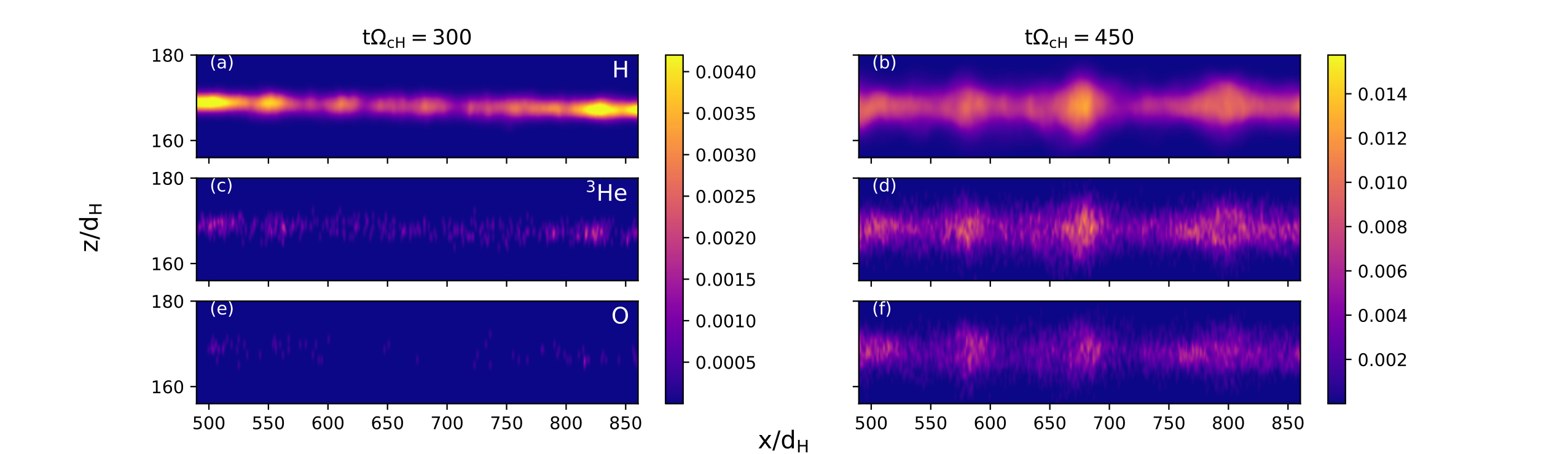}
\caption{%\textbf{Different ion species are magnetized and accelerated at different times.} 
The energetic particles density above injection energy (H - top, \ce{^3He} - middle, O - bottom) in the $x-z$ plane (averaged over $y$) at two times.\label{fig3}}
\end{figure*}

The delayed onset of Fermi acceleration for lower Q/M ions is caused by their larger gyroradii after injection: they get magnetized at later times when flux ropes and their adjacent exhausts grow large enough. We demonstrate this in Figure \ref{fig3} with the density of several ion species (normalized by their initial density) beyond their injection energies   (Equation \ref{inj_en}) at different times (grey lines in Figure \ref{fig2}(c)), around the region filled with flux ropes. At $t\Omega_{cH}=300$, with relatively small flux ropes ($\sim5d_H$ in $z$), protons (post-injection gyroradius $\rho_H\sim1.4d_H$ taking $\varepsilon_{inj}\sim1m_HV_A^2$) have already started Fermi acceleration for some time with many particles beyond the injection energy, while \ce{^3He} just started, and most Oxygens (post-injection $\rho_O\sim3.2d_H$) are not accelerated. At $t\Omega_{cH}=450$ the flux ropes become larger ($\sim10d_H$ in $z$) and all ion species are magnetized after injection, allowing continuous Fermi acceleration. Note that there is a short-term $\varepsilon_{max}$ increase before $t\Omega_{ci}=300$ (Figure \ref{fig2}(c)) for all species distinct from Fermi acceleration, more apparent for heavy ions. This is because a far-downstream portion (hundreds of $d_H$ from the x-line) of the large exhausts at this early time reaches a somewhat higher exhaust speed ($\sim0.8V_A$). At later time after the exhausts break up into flux ropes, this effect vanishes and gets overwhelmed by Fermi acceleration. 

To further elucidate this mechanism, we perform the following scaling analysis. The gyroradius after injection {\color{black}$\rho_x\propto (Q_x/M_x)^{-1}\sqrt{\varepsilon_{inj,x}}$ for a species $x$, with $\varepsilon_{inj,x}$ given in Equation \ref{inj_en}}. Assuming flux ropes grow linearly over time, the starting time of magnetization and acceleration {\color{black}$t_0\propto (Q_x/M_x)^{-1}\sqrt{\varepsilon_{inj,x}}$}. 
During Fermi acceleration we set
\begin{equation}
    \varepsilon_{max,x}\sim C_xt^\gamma \label{Emax_t},
\end{equation}
where $C_x$ is a species-specific constant. Since $\varepsilon_{max}=\varepsilon_{inj}$ at $t=t_0$, we obtain 
{\color{black}\begin{equation}\label{Emax_ratio}
    \varepsilon_{max,x}/\varepsilon_{max,H}\propto C_x/C_H\propto (Q_x/M_x)^\gamma (\frac{\varepsilon_{inj,x}}{\varepsilon_{inj,H}})^{1-\gamma/2}.
\end{equation}
When $\beta\ll1$, $(\frac{\varepsilon_{inj,x}}{\varepsilon_{inj,H}})^{1-\gamma/2}\sim 1$}.
Since the fitted scaling $\varepsilon_{max,x}/\varepsilon_{max,H}\propto (Q_x/M_x)^\alpha$, we have $\alpha$ related to the acceleration exponent: $\alpha\sim \gamma.$
% \begin{equation}
%     \alpha\sim \gamma.
% \end{equation}
In our low-$\beta$ simulations, $\gamma\sim 0.75$ (Figure \ref{fig2}(c)) roughly agrees with $\alpha\sim 0.6$ (Figure \ref{fig2}(b)). A higher initial $\beta$ approaching unity will introduce corrections {\color{black}since $(\frac{\varepsilon_{inj,x}}{\varepsilon_{inj,H}})^{1-\gamma/2}< 1$ in Equation \ref{Emax_ratio}}, increasing the relative difference of $\varepsilon_{max}$ and therefore $\alpha$. {\color{black}Since $\gamma$ may also change with $\beta$, a more detailed understanding will require future study. } Note that Equation \ref{Emax_t} implies the maximum gyroradius scales in time as $\rho_{max,x}\propto\sqrt{\varepsilon_{max,x}}\propto t^{\gamma/2}\sim t^{0.38}$, which grows much slower than flux ropes ($\propto t$), enabling the highest energy particles to stay magnetized during Fermi acceleration.

\textbf{Discussion and Conclusion.---}
%{In intermediate guide field regimes ($b_g\sim0.2-0.4$), electron pressure anisotropy may develop due to parallel electric fields to approach the firehose-instability limit (but not exceeding it) at the center of reconnection exhaust (shown by 2D fully kinetic simulations \cite{Le2013, Egedal2012, Egedal2015}). This could affect the field line tension locally, and lead to an elongated current layer (with electron-scale thickness) embedded at the exhaust center \cite{Le2013}. However, 2D kinetic simulations with parameters that capture these effects show little difference for ion acceleration from simulations that do not capture them \cite{Li2019a}. Also, such electron anisotropy may be further scattered/isotropized by small-scale structures in 3D reconnection-driven fluctuations \cite{Le2019, Li2019b}.} 
In summary, our 3D hybrid simulations demonstrate simultaneous nonthermal acceleration of all available ion species (up to {\ce{^{56}Fe^{14+}}}) in magnetic reconnection. We have uncovered the 3D turbulent dynamics and the fundamental mechanisms of particle injection and acceleration for multi-species ion acceleration, with strong implications to not only heliophysics but also astrophysics -- such as stellar flares and accretion-disk flares \citep{Ripperda2020,Nathanail2020} with nonrelativistic/transrelativistic magnetization. {\color{black}In a real system with open boundaries and escape of magnetic flux, reconnection can keep occurring to produce new flux ropes, so the dynamics in our simulations can occur repeatedly, rather than being transient.}

Our hybrid simulations only take into account the electron adiabatic heating without electron acceleration and pressure anisotropy, which can potentially influence magnetic tension and energy release. We also neglect potential electron-driven instabilities that affect ion acceleration, e.g. \citep{Fisk1978ApJ}, which needs further studies. However, previous studies suggest that electrons have less energy gain and pressure anisotropy than protons in reconnection \citep{Phan2013,Phan2014,Zhang2019,Zhang2019b,Shay2014,Haggerty2015,Zhang2021} due to weaker gain from Fermi reflection. {\color{black}In the Appendix, hybrid simulations indeed produce proton spectra similar to fully kinetic simulations \citep{Zhang2021}.} 
%{#cite appendix here and before} Our hybrid simulations also produce exponents ($p\sim4,\gamma\sim0.8$) for protons at low $\beta$ close to fully kinetic simulations \citep{Zhang2021}, further demonstrating that the hybrid model is viable for studying ion acceleration. 

Our predicted spectrum features ($p,\alpha$) may naturally account for the current observations near HCS and the magnetotail with low guide fields. { While the HCS observation \citep{Desai2022} (from PSP encounter 7) has a peak ion intensity occurring just outside the reconnection exhaust, this is not a common feature for HCS crossings. More recent PSP encounters such as 10 and 11 have found peak intensities inside the exhausts \cite{Desai2022AGU}. The observed profiles are likely affected by not only acceleration but also transport, and therefore may be highly variable across encounters. Detailed comparisons with simulations will require a future statistical study with many crossings beyond the scope of this Letter.} The remote sources like interchange reconnection and solar flares need to be further explored, where parameters are much less constrained, e.g. \citep{Dahlin2022ApJ}. The 3D flux-rope dynamics and the dependence of features (such as $p,\alpha, \varepsilon_{inj}$) on parameters can be compared in details with future spacecraft measurements, which is critical for understanding particle acceleration in reconnection.

\begin{acknowledgements}
\textbf{Acknowledgment.---}
We gratefully acknowledge the helpful discussions in the SolFER DRIVE Science Center collaboration. We also acknowledge technical support from Xiaocan Li at Dartmouth College, and Adam Stanier at Los Alamos National Laboratory. Q.Z, F.G., W.D. and H.L. acknowledge the support from Los Alamos National Laboratory through the LDRD program and its Center
for Space and Earth Science (CSES), DOE OFES, and NASA programs through grant NNH17AE68I, 80HQTR20T0073, 80NSSC20K0627, 80HQTR21T0103 and 80HQTR21T0005, and through Astrophysical Theory Program. 
The simulations used resources provided by the Los Alamos National Laboratory Institutional Computing Program, the National Energy Research Scientific Computing Center (NERSC) and the Texas Advanced Computing Center (TACC). 
\end{acknowledgements}

\setcounter{figure}{0}
\makeatletter 
\renewcommand{\thefigure}{A\@arabic\c@figure}
\makeatother

\begin{figure*}
\includegraphics[width=1\textwidth]{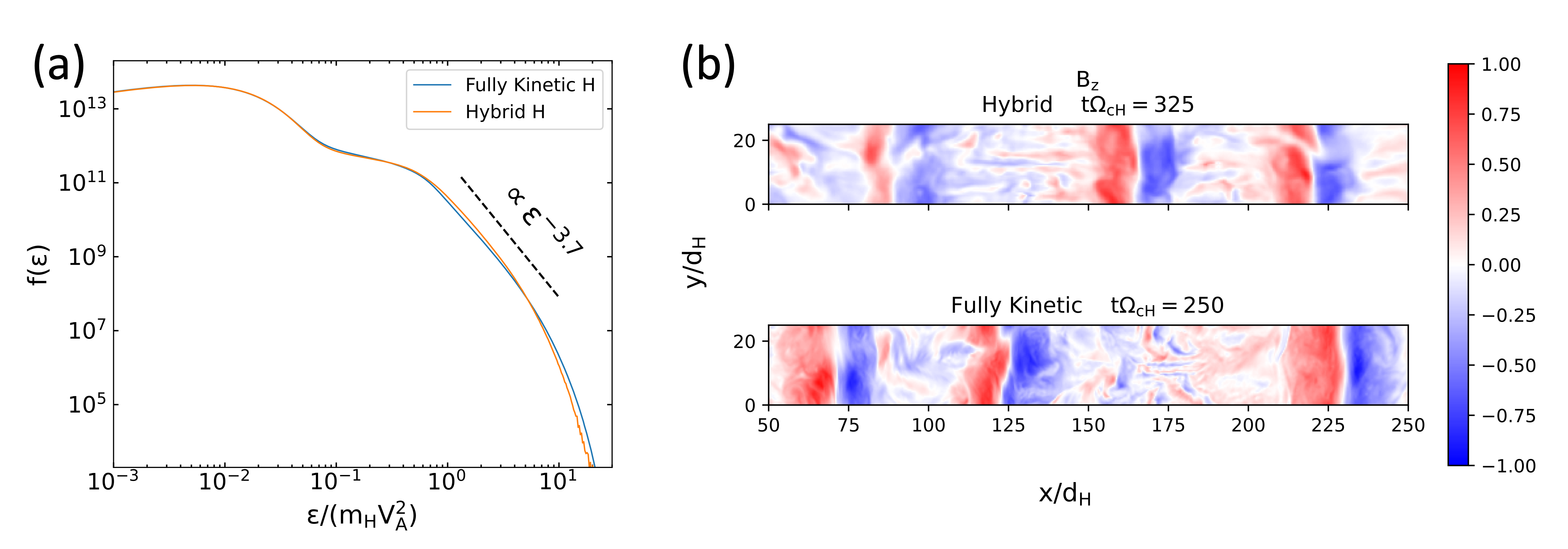}
\caption{
{\color{black}Comparison of hybrid and fully kinetic simulations with parameters $L_x=300d_H$, $L_y=25d_H$, $b_g=0.2$, $T=0.01m_HV_A^2$. (a) Proton spectra (normalized to have the same total number) when hybrid ($t\Omega_{cH}=475$) and fully kinetic ($t\Omega_{cH}=400$) runs have similar reconnected magnetic flux. (b) Comparison of magnetic flux ropes by showing Bz on a x-y cut in the middle of the reconnection layer at simulation times when hybrid ($t\Omega_{cH}=325$) and fully kinetic ($t\Omega_{cH}=250$) runs have similar reconnected magnetic flux.}\label{figA.1}  }
\end{figure*}

{\color{black}
\textbf{Appendix on Comparing Hybrid and Fully Kinetic Simulations.---}
We show here a direct comparison of the hybrid and the fully kinetic \cite{Zhang2021} simulations, and their results are very similar. We use the same physical parameters in \cite{Zhang2021} for the hybrid simulation: $L_x=300d_H$, $L_y=25d_H$, $b_g=0.2$, $T=0.01m_HV_A^2$, with only the proton ion species. Due to the double periodic setup for the hybrid code, we use $L_z=250d_H$ for the hybrid run, which doubles $L_z=125d_H$ of the fully kinetic run. We compare snapshots from both runs at times with similar reconnected magnetic flux. As shown in Figure \ref{figA.1}, the proton energy spectra for both runs closely resemble each other. As for the flux rope dynamics, both runs have kink unstable flux ropes but they are more coherent (not fragmented) due to the small domain size compared to those with large domains in the Figure \ref{fig1}. }
% \bibliography{reference}
%\bibliographystyle{aasjournal}

% \begin{thebibliography}{41}%

% \end{thebibliography}%

% \bigskip
% \bigskip
% \bigskip

% \clearpage
% \newpage
% \pagebreak

\section{supplemental material}

See Figure \ref{figE.1} - Figure \ref{figE.5}.

\setcounter{figure}{0}
\makeatletter 
\renewcommand{\thefigure}{S\@arabic\c@figure}
\makeatother
\begin{figure*}
\includegraphics[width=0.9\textwidth]{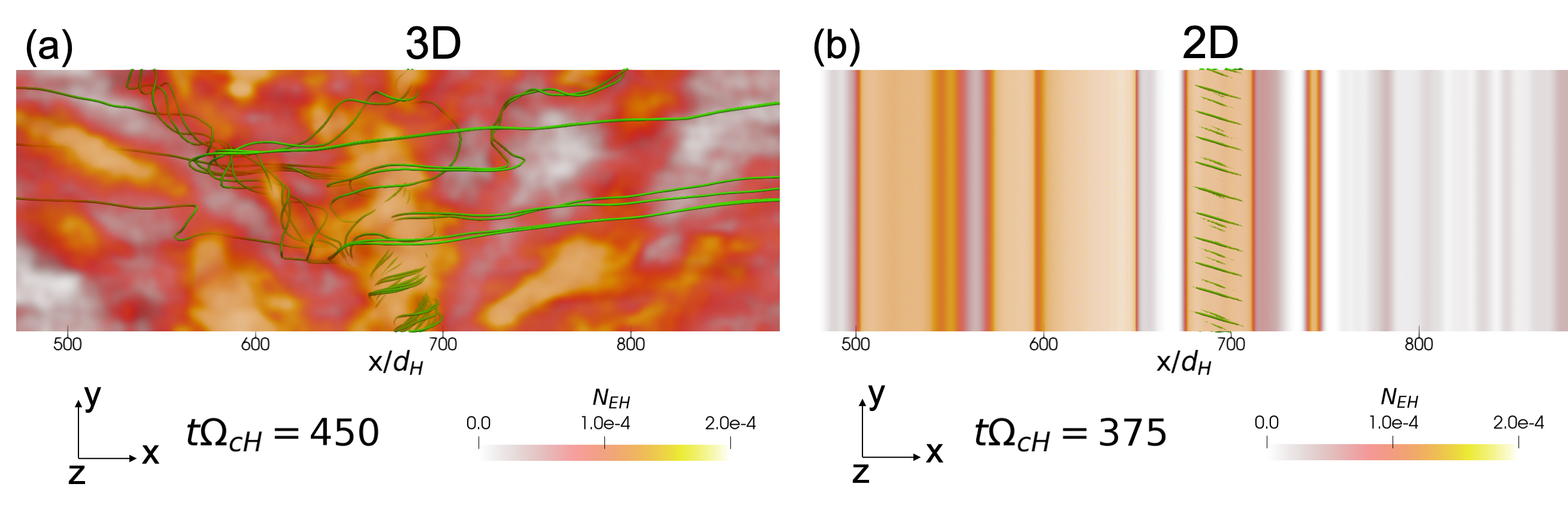}
\caption{\textbf{3D field-line chaos detraps particles from flux ropes.} (a) energetic proton density $N_{EH}$ in the energy range [3.4$m_HV_A^2$,6.8$m_HV_A^2$] at $t\Omega_{cH}=450$ in the 3D simulation with $\beta_H=0.18$. This has the same perspective as Figure 1(b). We also show 17 sample field lines starting from the core of a flux rope at x=690 near the bottom. The field lines become chaotic and connect outside of the flux rope and also to other fragmented flux ropes, enabling particles to spread throughout the reconnection layer for efficient acceleration. (b) in comparison, a similar demonstration for the 2D counterpart without variation in $y$ at $t\Omega_{cH}=375$ with an similar amount of magnetic flux reconnected as panel (a). The field lines and energetic particles are mostly confined within the flux rope, hindering efficient acceleration.   \label{figE.1}}
\end{figure*}

\begin{figure*}
\includegraphics[width=0.9\textwidth]{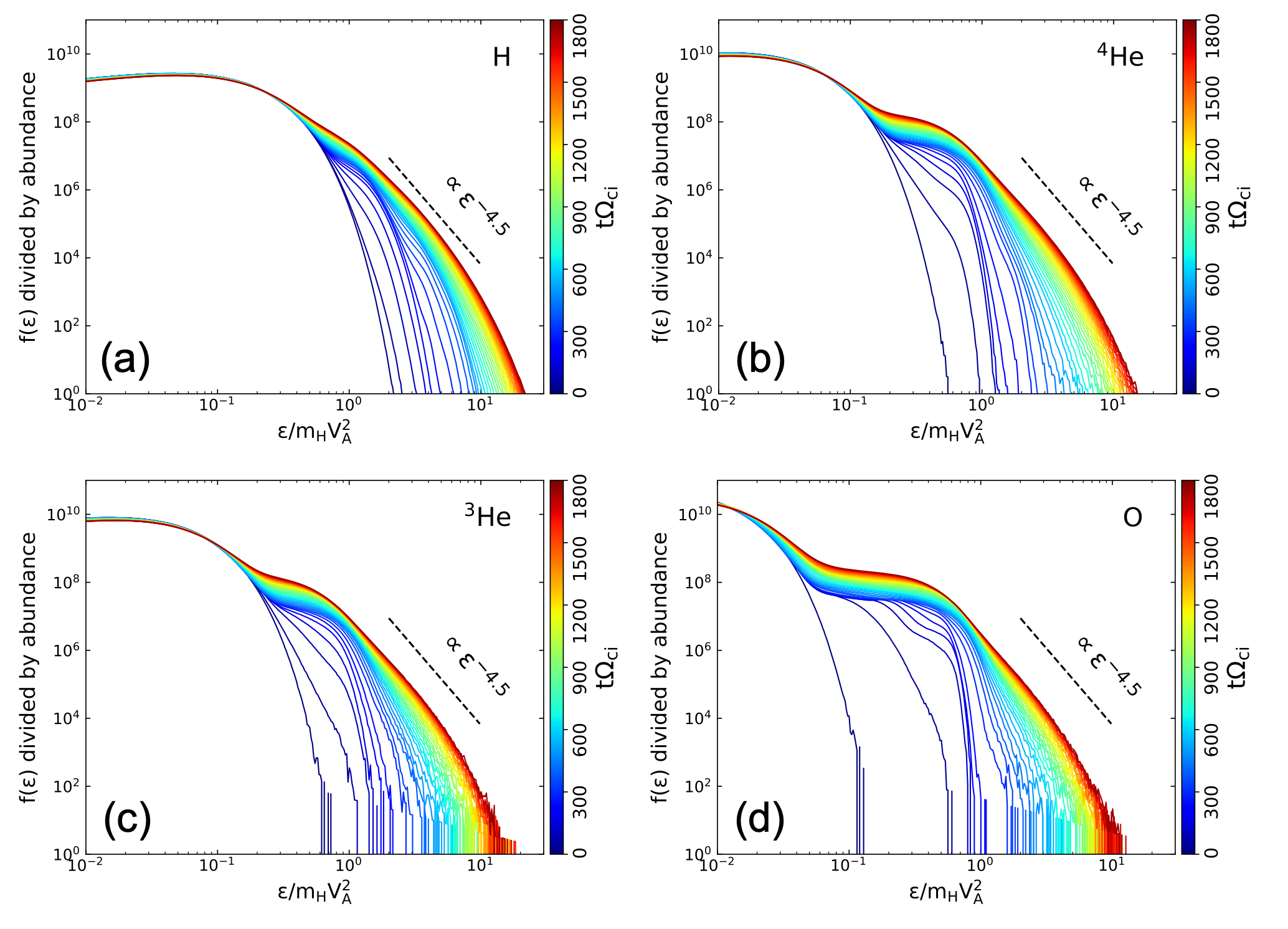} %figE.2.png}
\caption{{\textbf{All available ion species form nonthermal power laws over time.} The time evolution of the ion energy spectra versus energy per nucleon $\varepsilon$ in the 3D simulation ($\beta_H=0.18$) for \ce{H, ^4He, ^3He and O}. The power laws extend along approximately the indicated spectral slope towards higher energy over time. }
\label{figE.3}}
\end{figure*}

\begin{figure*}
\includegraphics[width=0.7\textwidth]{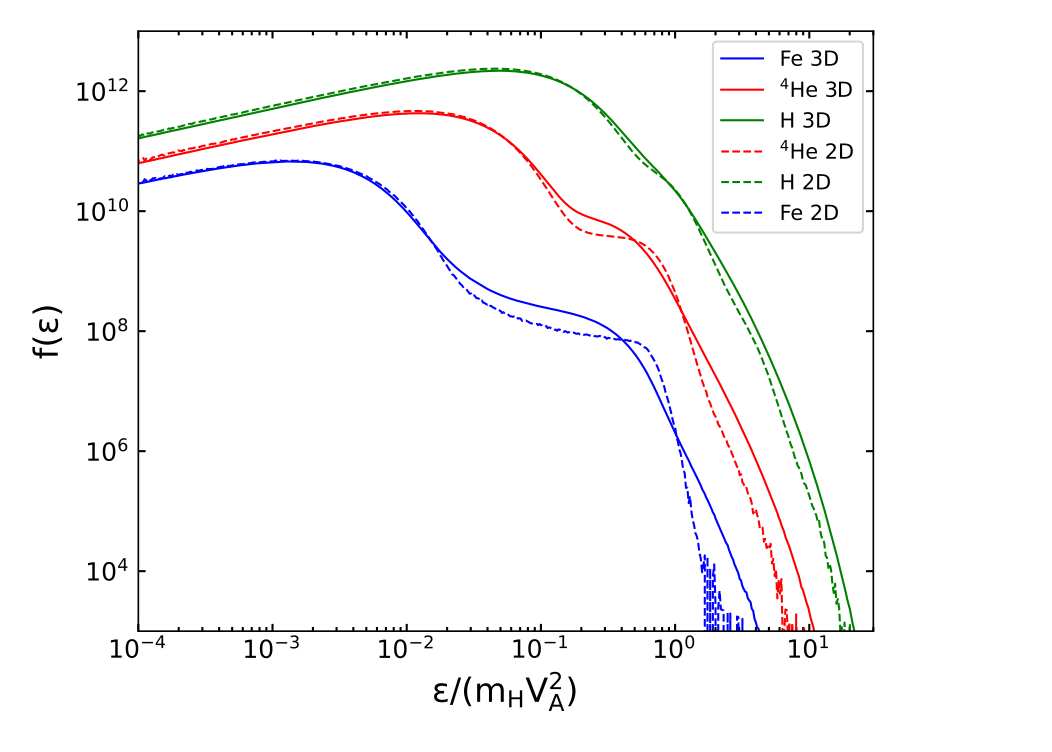} %figE.2.png}
\caption{\textbf{3D simulations have more efficient acceleration than 2D for all ion species.} Ion energy spectra versus energy per nucleon $\varepsilon$ in the 3D simulation ($\beta_H=0.18$) at $t\Omega_{cH}=1800$ in comparison to the 2D counterpart at the time when a similar amount of magnetic flux is reconnected. The 2D simulation are normalized to have the same number of particles as 3D. The spectra of different species are not normalized by abundance (unlike Figure 2(a)) so major and minor ions are separated in scales from each other. \label{figE.2}}
\end{figure*}

% \begin{figure*}
% \includegraphics[width=1.0\textwidth]{figE.4.png} %figE.2.png}
% \caption{{\textbf{The Fermi process is the dominant acceleration mechanism for all ion species.} The particle acceleration rate as a function of energy per nucleon $\varepsilon$ for several major acceleration mechanisms (as different particle drifts along electric fields in the guiding center approximation in Dahlin et al. 2014 \citep{Dahlin2014} and Li et al. 2017 \citep{Li2017}), calculated for \ce{H, ^4He, and ^3He} at $t\Omega_{cH}=900$. We show mechanisms of curvature drift (representing the Fermi acceleration), gradient drift, parallel electric field, as well as the total acceleration. The Fermi acceleration is the dominant acceleration mechanism for the high energy ions for each species. The parallel electric field acceleration is negligible for ions and the gradient drift term is an energy sink due to the magnetic-moment conservation and the overall reduction of magnetic fields in reconnection. The minor ions, due to small abundance and less particles of high energy per nucleon, have larger numerical fluctuations at higher energies due to lack of simulation particles. However, it is clear that Fermi acceleration is the dominant process in our simulations.} \label{figE.4}}
% \end{figure*}
\begin{figure*}
\includegraphics[width=1\textwidth]{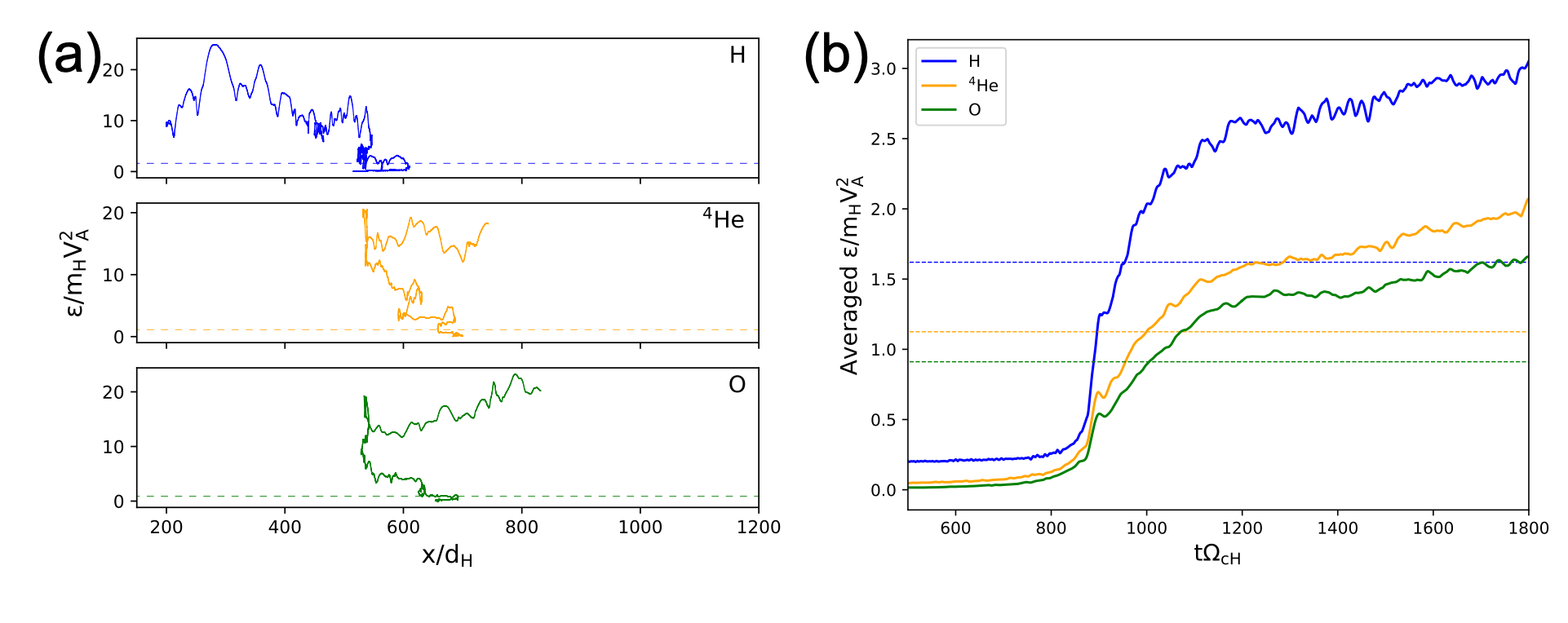} %figE.2.png}
\caption{
\textbf{All ion species can be injected by a single Fermi reflection at an exhaust before further Fermi acceleration.}
(a) single-particle trajectories for three species \ce{H, ^4He and O}. As soon as the particle first crosses an exhaust, it gets kicked by the exhaust outflow through a single Fermi reflection to reach the order of its injection energy per nucleon $\varepsilon_{inj}$ (the dash horizontal line for each species), then it can travel around getting Fermi bounces for further acceleration. (b) for a further statistical study, we average the energy per nucleon of particles injected around $t\Omega_{cH}=900$. We observe a two-phase acceleration trend: different species get injected quickly to above their specific injection energy per nucleon (dash lines, through the Fermi reflection shown above), and then they start the prolonged Fermi acceleration. \label{figE.5}}
\end{figure*}

\end{document}